\documentclass[conference]{IEEEtran}
\usepackage{fancyhdr}
\IEEEoverridecommandlockouts
\usepackage{cite}
\usepackage{amsmath,amssymb,amsfonts}
\usepackage{iftex}
\iftutex
\usepackage{fontsetup}
\fi
\usepackage{graphicx}
\usepackage{textcomp}
\usepackage{xcolor}
\def\BibTeX{{\rm B\kern-.05em{\sc i\kern-.025em b}\kern-.08em
    T\kern-.1667em\lower.7ex\hbox{E}\kern-.125emX}}

\usepackage{tikz}
\usetikzlibrary{positioning}

\usepackage{array}
\usepackage{booktabs}
\usepackage{adjustbox}
\usepackage{tabularx}
\usepackage{physics}
\usepackage[inline]{enumitem}
\usepackage{algorithm}
\usepackage{algpseudocode}
\usepackage{subfigure}
\usepackage{amsthm}
\usepackage{amsmath}
\usepackage{placeins}
\usepackage{url}
\usepackage{mdframed}
\usepackage{multirow}
\usepackage{flushend}

\newlist{lenum}{enumerate}{1}
\setlist[lenum,1]{label=(L\arabic*)}

\usepackage{calc} 
\newcolumntype{L}[1]{>{\raggedright\let\newline\\\arraybackslash\hspace{0pt}}m{#1}}
\newcolumntype{C}[1]{>{\centering\let\newline\\\arraybackslash\hspace{0pt}}m{#1}}
\newcolumntype{R}[1]{>{\raggedleft\let\newline\\\arraybackslash\hspace{0pt}}m{#1}}

\newtheoremstyle{boldfinding} 
    {3pt}
    {3pt}
    {\itshape}
    {}
    {\bfseries}
    {.}
    {.5em}
    {}

\theoremstyle{boldfinding}
\newtheorem{finding}{Finding}

\usepackage[acronym]{glossaries}
\newacronym{API}{API}{Application Programming Interface}
\newacronym{AR}{AR}{augmented reality}
\newacronym{BIM}{BIM}{Building Information Modeling}
\newacronym{CDU}{CDU}{cooling distribution unit}
\newacronym{CDUP}{CDUP}{\gls{CDU} Pump}
\newacronym{CEP}{CEP}{Central Energy Plant}
\newacronym{CT}{CT}{cooling tower}
\newacronym{CTW}{CTW}{cold temperature water}
\newacronym{CTWP}{CTWP}{cold temperature water pump}
\newacronym{CTWS}{CTWS}{cold temperature water supply}
\newacronym{CTWR}{CTWR}{cold temperature water return}
\newacronym{DT}{DT}{digital twin}
\newacronym{exadigit}{ExaDigiT}{{\color{Red}WRONG USAGE}}
\glsunset{exadigit} 
\newacronym{EDTF}{EDT-F}{\gls{EDT}-Frontier}
\newacronym{EHX}{EHX}{intermediate heat exchanger}
\newacronym{FMI}{FMI}{Functional Mock-up Interface}
\newacronym{FMU}{FMU}{Functional Mock-up Unit}
\newacronym{GPM}{GPM}{Gallons Per Minute}
\newacronym{HPL}{HPL}{High Performance Linpack}
\newacronym{HTW}{HTW}{hot temperature water}
\newacronym{HTWP}{HTWP}{hot temperature water pump}
\newacronym{HTWS}{HTWS}{hot temperature water supply}
\newacronym{HTWR}{HTWR}{hot temperature water return}
\newacronym{IDE}{IDE}{integrated development environment}
\newacronym{MAE}{MAE}{mean absolute error}
\newacronym{ODE}{ODE}{ordinary differential equation}
\newacronym{ORNL}{ORNL}{Oak Ridge National Laboratory}
\newacronym{OLCF}{OLCF}{Oak Ridge Leadership Computing Facility}
\newacronym{PID}{PID}{proportional-integral-derivative}
\newacronym{PUE}{PUE}{Power Usage Effectiveness}
\newacronym{RAPS}{RAPS}{Resource Allocator and Power Simulator}
\newacronym{RMSE}{RMSE}{root mean square error}
\newacronym{SOTA}{SOTA}{state-of-the-art}
\newacronym{SIVOC}{SIVOC}{Super Intermediate Voltage Converter}
\newacronym{trans}{TRANSFORM}{Transient Simulation Framework of Reconfigurable Models}
\newacronym{UEFMI}{UEFMI}{Unreal Engine Functional Mockup Interface Plugin}
\newacronym{UE5}{UE5}{Unreal Engine 5}
\newacronym{VR}{VR}{virtual reality}
\newacronym{HMD}{HMD}{Head Mounted Display}

\glsdisablehyper

\begin{document}

\title{A Digital Twin Framework for Liquid-cooled Supercomputers as Demonstrated at Exascale  \\
\thanks{Notice: This manuscript has been authored by UT-Battelle, LLC, under contract DE-AC05-00OR22725 with the US Department of Energy (DOE). The US government retains and the publisher, by accepting the article for publication, acknowledges that the US government retains a nonexclusive, paid-up, irrevocable, worldwide license to publish or reproduce the published form of this manuscript, or allow others to do so, for US government purposes.}}

\makeatletter
\newcommand{\linebreakand}{
  \end{@IEEEauthorhalign}
  \hfill\mbox{}\par
  \mbox{}\hfill\begin{@IEEEauthorhalign}
}
\makeatother

\author{\IEEEauthorblockN{Wesley Brewer}
\IEEEauthorblockA{\textit{Oak Ridge National Laboratory} \\
Oak Ridge, TN, USA \\
brewerwh@ornl.gov}
\and
\IEEEauthorblockN{Matthias Maiterth}
\IEEEauthorblockA{\textit{Oak Ridge National Laboratory} \\
Oak Ridge, TN, USA \\
maitherthm@ornl.gov}
\and
\IEEEauthorblockN{Vineet Kumar}
\IEEEauthorblockA{\textit{Oak Ridge National Laboratory} \\
Oak Ridge, TN, USA \\
kumarv@ornl.gov}

\linebreakand

\IEEEauthorblockN{Rafal Wojda}
\IEEEauthorblockA{\textit{Oak Ridge National Laboratory} \\
Oak Ridge, TN, USA \\
wojdarp@ornl.gov}
\and
\IEEEauthorblockN{Sedrick Bouknight}
\IEEEauthorblockA{\textit{Oak Ridge National Laboratory} \\
Oak Ridge, TN, USA \\
bouknightsl@ornl.gov}
\and
\IEEEauthorblockN{Jesse Hines}
\IEEEauthorblockA{\textit{Oak Ridge National Laboratory} \\
Oak Ridge, TN, USA \\
hinesjr@ornl.gov}

\linebreakand

\IEEEauthorblockN{Woong Shin}
\IEEEauthorblockA{\textit{Oak Ridge National Laboratory} \\
Oak Ridge, TN, USA \\
shinw@ornl.gov}
\and
\IEEEauthorblockN{Scott Greenwood}
\IEEEauthorblockA{\textit{Oak Ridge National Laboratory} \\
Oak Ridge, TN, USA \\
greenwoodms@ornl.gov}
\and
\IEEEauthorblockN{David Grant}
\IEEEauthorblockA{\textit{Oak Ridge National Laboratory} \\
Oak Ridge, TN, USA \\
grantdr@ornl.gov}
\linebreakand

\IEEEauthorblockN{Wesley Williams}
\IEEEauthorblockA{\textit{Oak Ridge National Laboratory} \\
Oak Ridge, TN, USA \\
williamswc@ornl.gov}
\and
\IEEEauthorblockN{Feiyi Wang}
\IEEEauthorblockA{\textit{Oak Ridge National Laboratory} \\
Oak Ridge, TN, USA \\
fwang2@ornl.gov}
}

\maketitle

\thispagestyle{fancy}
\lhead{}
\rhead{}
\chead{}
\rfoot{}
\cfoot{}
\renewcommand{\headrulewidth}{0pt}
\renewcommand{\footrulewidth}{0pt}

\begin{abstract}
We present \gls{exadigit}, an open-source framework for developing comprehensive digital twins of liquid-cooled supercomputers. It integrates three main modules: (1) a resource allocator and power simulator, (2) a transient thermo-fluidic cooling model, and (3) an augmented reality model of the supercomputer and central energy plant. The framework enables the study of ``what-if'' scenarios, system optimizations, and virtual prototyping of future systems. Using Frontier as a case study, we demonstrate the framework's capabilities by replaying six months of system telemetry for systematic verification and validation. Such a comprehensive analysis of a liquid-cooled exascale supercomputer is the first of its kind. \gls{exadigit} elucidates complex transient cooling system dynamics, runs synthetic or real workloads, and predicts energy losses due to rectification and voltage conversion. Throughout our paper, we present lessons learned to benefit HPC practitioners developing similar digital twins. We envision the digital twin will be a key enabler for sustainable, energy-efficient supercomputing. 
\end{abstract}


\section{Introduction}

A drastic reduction in power consumption was the key to realizing exascale supercomputing. In 2008, DARPA published a study where they made projections about what it would take to reach a target design for exascale computing of 20 megawatts (MW) per exaflop (EF) \cite{bergman2008exascale}. They projected that an exascale system could be achieved in 2015, requiring between 68 and 155 MW/EF \cite{atchley2023frontier}. Frontier was deployed in 2021 and achieved 1.102 EF in June 2022 using an average power of 21.1 MW \cite{Top500_2022}; its performance improved in November 2023 to 1.194 EF performance at 22.7 MW \cite{Top500_2023}, and again in June 2024 to 1.206 EF at 22.8 MW \cite{Top500_2024}. To connect the dots on how this has played out over the past 15 years, consider that in 2009 technology the projected energy cost of scaling the Jaguar supercomputer to exascale performance would have required about three gigawatts. This was markedly reduced down to 330 MW/EF with the advent of GPUs in the Titan supercomputer in 2012, and further down to 65 MW/EF for Summit in 2018, and finally down to 19 MW/EF for Frontier. At the same time each generation of supercomputer has achieved a tenfold increase in performance: from 2.5 petaflops (PF) for Jaguar, to 27 PF for Titan, to 200 PF for Summit, to 2 EF for Frontier. 

While significant efforts in hardware optimization have driven these drastic advancements in efficiency, we have reached a point where the system is so highly optimized that little room remains for further improvements. 
On the operational side, two main areas of research have been instrumental for improving datacenter efficiency: simulations \cite{zuo2021improving}, and analysis of system telemetry \cite{shin2021revealing}. 
Additional improvements necessitate innovative tools that focus on end-to-end improvement, such as mixed-precision iterative refinement \cite{haidar2020mixed}, AI-based surrogate models \cite{su20231}, and \glspl{DT}. \textit{We hypothesize that developing a comprehensive digital twin of both the facility as well as the supercomputer will provide a robust tool for end-to-end optimization across multiple facets.}

\Glspl{DT} have emerged as a means of merging both telemetry and simulations to develop a holistic virtual representation of the system, bridging both the physical and virtual worlds. The AIAA digital engineering integration committee defines a \gls{DT} as \cite{aiaa2020digital}:

\begin{quote}
“...a set of virtual information constructs that mimics the structure, context, and behavior of an individual / unique physical asset, or a group of physical assets, 
is dynamically updated with data from its physical twin throughout its life cycle, and 
informs decisions that realize value.”
\end{quote}

\noindent Here, the mimicking \textit{structure} refers to the 3D modeling of the physical assets (racks, servers, pumps, etc.), while mimicking \textit{behavior} refers to developing either simulations or AI/ML models, and \textit{dynamically updated} speaks of telemetry data generated from the physical twin. 

\Glspl{DT} provide value proposition at multiple stages of the data center lifecycle: (1) planning/design, (2) construction, and (3) operations \cite{venturebeat_digital_twins_2023}. During the planning/design phase, the \gls{DT} can enable more informed decisions for future acquisitions, and provide predictive capabilities of energy efficiency for future systems. This can be accomplished via virtual prototyping capabilities, such as the ability to virtually design and test the cooling system, or design virtual networks to study multi-application interactions on congestion performance. During the construction phase, the \gls{DT} streamlines construction costs, reduces construction waste, and reduces outages when retrofitting systems. During the operations phase, the \gls{DT} enables predictive maintenance, extends asset life through reliability and availability modeling, and enhances cooling efficiency by optimizing system behavior. For these reasons and more, digital twins are widely becoming ``management best practice" in various industries \cite{violino2023}.

To build such a tool, we needed to be able to integrate both system telemetry and simulation, integrate models from across multiple domains, and demonstrate the approach through real-world examples. To that end, in this paper we provide the following specific contributions:

\begin{enumerate}
    \item An open-source reusable digital twin framework for data centers, \gls{exadigit}, with modular sub-components supported by telemetry and simulation, 
    \item Extensive verification and validation studies of the framework,
    \item Demonstration of the framework at exascale for Frontier.
\end{enumerate}

Our paper is structured as follows: In Section \ref{sec:background}, we first give a background of other research related to digital twin modeling of the various facets of the data center, then in Section \ref{sec:architecture} we give an overview of the \gls{exadigit} architecture, discussing how each component was developed, then in Section \ref{sec:vv} we discuss verification and validation of the framework, and finally close with Section \ref{sec:conclusions} on conclusions and future work. Throughout the various sections, we have highlighted valuable lessons learned, denoted as ``findings". 

\section{Background}
\label{sec:background}

Research towards development of digital twins for data centers has mainly been separate, siloed efforts focused on either data center cooling, network performance, power consumption, or visualization efforts. We discuss the history of research for energy consumption, cooling models, and visual analytics, identifying the \gls{SOTA} techniques in each area, highlighting for each area how our work differs. 

\subsubsection{Energy Consumption Models} 
There have been a couple of rather extensive surveys of previous work on power consumption modeling of \emph{data centers}, which are generally air-cooled CPU-only systems. Dayarathna and Wen \cite{dayarathna2015data} conducted an in-depth analysis of over 200 power consumption models. From their detailed analysis, they deduced that as of 2015: (1) a significant portion of the research primarily focused on the lower levels of the data center hierarchy, often overlooking the broader perspective of the entire facility, (2) many studies considered only a limited set of server metrics, (3) the accuracy of these models remained a major concern. More recently, Jin et al. \cite{jin2020review} surveyed literature on power consumption modeling efforts of CPU-only air-cooled servers. They analyzed 47 different models, which they classify into the following model categories: additive, BA (baseline power + active power), simple regression, multiple regression, power function, non-linear, polynomial, and other. Regarding GPU-enabled HPC systems, Sîrbu and Babaoglu's work \cite{sirbu2018data} represents the \gls{SOTA} in power systems modeling of HPC systems. They developed a predictive model for total system power of a hybrid CPU-GPU-MIC supercomputer. Their model is based on three components: (1) using support vector regression to predict power per job before the jobs are started, (2) predicting job duration via a simple heuristic, and (3) predicting total system power based on the measured power of computing units. They mention that few models address system-level prediction of power, a point that was originally made by Dayarathna and Wen \cite{dayarathna2015data}. Our model focuses on predicting the total system power at fixed time intervals as well as modeling the losses due to energy conversion. 

\begin{figure*}[t]
    \centering
    \includegraphics[width=5in]{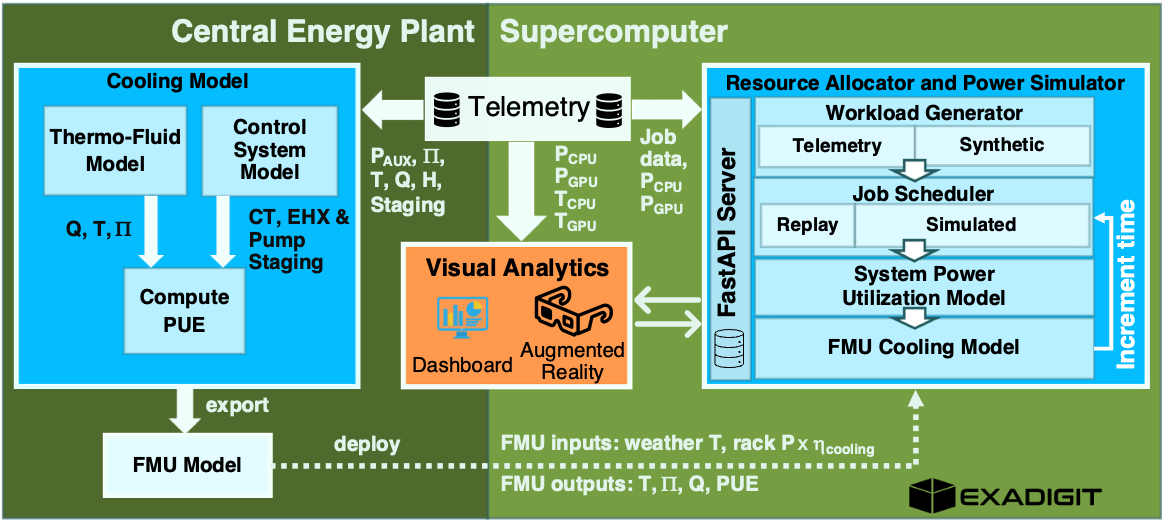}
    \caption{\gls{exadigit} architectural overview.}
    \label{fig:arch}
\end{figure*}

\subsubsection{Thermo-Fluid Cooling Models} 
Most of the work on data center cooling has been focused on air-cooled systems, e.g. \cite{lee2013analysis, ham2016impact, fu2018modelica}, especially focused on \emph{cooling efficiency}. Zohdi \cite{zohdi2022digital} presents the development of a digital twin and machine learning framework to model an idealized air-cooled system with thermal effects using direct numerical simulation (DNS). The machine learning framework uses genetic algorithms to learn the optimal parameters, such as flow rates in/out of multiple room vents. Moreover, Zhang et al. \cite{zhang2022smart} built a digital twin for data centers called ``Smart DC''. Their framework uses a combination of Computational Fluid Dynamics (CFD), 6SigmaDC \cite{6sigmaDC}, demonstrated for an air-cooled data center, along with AI in the form of XGBoost to optimize the control parameters of the air conditioning system in order to optimize the \gls{PUE} of a datacenter. They show that their method can effectively reduce the \gls{PUE} from 1.15 to 1.08. In terms of liquid-cooled systems, Heydari et al. \cite{heydari2022liquid} represents the \gls{SOTA} in this field, having performed extensive analysis of liquid-cooled systems by running CFD simulations using both Macroflow \cite{macroflow} for blade-level flow loops, 6SigmaET \cite{6sigmaET} for thermal simulations of cold plates, along with 6SigmaRoom \cite{6sigmaDC} for simulating the air flow in the room. Whereas their cooling model is based on expensive proprietary software, our cooling model is built on an open-source Modelica framework \cite{modelica, greenwood2017a} and supports modeling the transient dynamics of the entire liquid cooling system. 

\subsubsection{Visual Analytics}
There has also been work primarily focused on the \emph{visualization} aspects of supercomputers, with efforts mainly targeting HPC monitoring \cite{rosenthal2018augmented,hpcmonitor4b,bergeron20213d}, visualizing file system activity \cite{averbukh2019visualizing}, and visualizing network traffic \cite{zhou2003graph,fujiwara2018visual,bhatele2016analyzing}. Bergeron et al. \cite{bergeron20213d} and Riha et al. \cite{hpcmonitor4b} represent the \gls{SOTA} in HPC monitoring. 
Our visual analytics approach is unique in that it combines telemetry and simulations in a single environment, supporting both \gls{AR} and dashboard, enabling the replay of live system telemetry or launching ``what-if'' simulation experiments. 

To summarize, there has been a significant amount of research investigating either power or cooling or visualizations, but a conspicuous lack of research attempting to build a comprehensive open-source tool for holistically modeling the system. The one exception that we are aware of is the work by NVIDIA \cite{NVIDIA2023}, which uses Cadence 6SigmaDCX and NVIDIA Modulus \cite{hennigh2021nvidia} to model an air-cooled data center, NVIDIA Air for modeling the network via Cumulus virtual machines, and NVIDIA Omniverse for managing and visualizing 3D assets. While we initially experimented with these tools, the toolkit did not provide support for modeling liquid-cooled systems, resource allocation, and energy consumption; therefore we decided to forge our own open-source framework. 

In this work, we present our development of an open-source digital twin framework, which is currently designed for modeling liquid-cooled systems. The open stack of software consists of a Python-based \gls{RAPS} module, a Modelica-based thermo-fluids cooling model, and C++-based 3D interactive augmented reality model utilizing \gls{UE5}, as shown in Fig. \ref{fig:arch}. 

\begin{figure*}[t]
    \centering
    \includegraphics[width=5in]{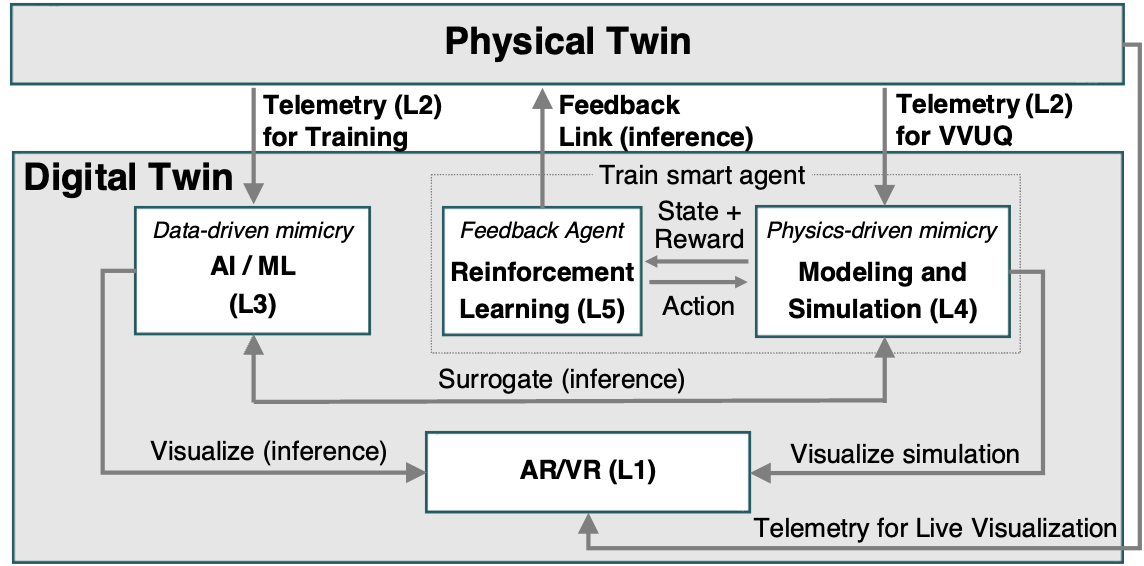}
    \caption{Relationships between \gls{DT} levels.}
    \label{fig:level_interactions}
\end{figure*}

\section{\gls{exadigit} Architecture}
\label{sec:architecture}

In this section we first give an overview of the high-level \gls{exadigit} architecture, and then step through the process of development from requirements analysis, technical specifications, and then discuss the details of the various components. Fig. \ref{fig:arch} shows the architectural overview of the various components of \gls{exadigit}. There are three main modules that we develop: (1) \gls{RAPS}, (2) a cooling model, and (3) visual analytics capabilities. 
The \gls{RAPS} module can replay workloads from telemetry, reschedule them, or simulate synthetic workloads on the supercomputer to analyze the resulting energy consumption; 
further details are provided in Section \ref{sec:raps}. The cooling model simulates thermo-fluid dynamics and control of the \gls{CEP}, which itself includes three components: (1) a thermo-fluid model for predicting temperatures ($T$), pressures ($\Pi$), and flow rates ($Q$); (2) a control system model for predicting the staging of cooling towers, hot/cold water pumps, and heat exchangers; and (3) a sub-module for predicting the system \gls{PUE}. This will be discussed in further detail in Section \ref{sec:cooling}. Both \gls{RAPS} and the cooling model can be interfaced either via a terminal console, the web-based dashboard, or the augmented reality environment for visual insights, which will be discussed in Section \ref{sec:visual_analytics} and shown in Fig. \ref{fig:interfaces}. 

Each module of the digital twin generally falls into one of the following five categories \cite{autodesk_digital_twins_2023}:

\begin{lenum}
    \item The \textit{descriptive} twin models the physical assets using both CAD models such as 3D modeling files (e.g., Autocad/Revit), as well as game engines (e.g., Unreal Engine, Unity, or NVIDIA Omniverse). 
    \item The \textit{informative} twin incorporates the telemetry data to provide real-time data insights into the physical twin. 
    \item The \textit{predictive} twin utilizes telemetry data to develop data-driven predictive models using AI/ML. 
    \item The \textit{comprehensive} twin leverages modeling and simulation techniques to provide virtual prototyping capabilities that are able to address ``what-if'' scenarios -- generally not in real-time. 
    \item The \textit{autonomous} twin uses techniques such as reinforcement learning to learn to make autonomous decisions for system optimization. 
\end{lenum}

\noindent These levels relate to each other as depicted in Fig. \ref{fig:level_interactions}. This paper covers using L1 for visualization, L2 for validation, and L4 for modeling and simulation. AI/ML L3 models are built on telemetry data, and therefore are fundamentally interpolative and thus often do not extrapolate well \cite{brunton2020machine}; however, they can generally be inferenced in real-time. L4 simulations, based on first principles, are the primary engine of digital twins \cite{zhao2023knowledge}; unlike AI/ML models, they generally require a longer development time, and utilize telemetry data for validation. They are also more computationally expensive, generally making real-time operation unfeasible. Despite these challenges, L4 simulations are extrapolative and can be effectively used for virtual prototyping. 
An alternative approach is to use the simulations to generate data to train a machine-learned \textit{surrogate model}, which has the advantage of being able to run in real-time, but can also be used to model virtual prototypes \cite{wang2023toward}. Reinforcement learning (L5) is used primarily for training autonomous agents that can be used to make control decisions in order to optimize processes. An example of L5 would be training an agent to perform automated setpoint control for improved cooling efficiency by minimizing setpoint overshoot of \gls{PID} controllers \cite{todd2021artificial}.  

\begin{finding}
Simulations are the primary building blocks of digital twins. However, machine-learned models should also have a significant role for modeling system workloads, i.e., application fingerprinting. Furthermore, system telemetry plays critical roles in our development of machine-learned models and for validation. 
\end{finding}

\subsection{Requirements Analysis and Gathering}

At the project's outset, we interviewed several HPC engineers to conduct a requirements analysis—our goal was to identify potential use cases for the digital twin and pinpoint opportunities for demonstrations.  We list some of the potential use cases here that resulted from those discussions:

\begin{itemize}
    \item Understanding dynamic energy consumption and losses in the system, as well as visualizing energy consumption on a per-job basis.  
    \item Virtually extending the cooling system to support a secondary HPC system in the future, and evaluating its impact on cooling performance of the current system.
    \item Understanding temperature problems in the past and problems with cooling loops by visualizing heat maps in the system.
    \item Early detection of thermal throttling--when the system has to reduce CPU or GPU clock frequency to reduce the heat  generated by the system. 
    \item Understanding how weather correlates to GPU temperatures on the system. 
    \item Understanding water quality issues and how they affect system performance. Water-based coolants can suffer from biological growth in the blade-level cooling system causing blockage to specific nodes. Can these types of blockages be detected?
    \item Studying effects of SCADA cybersecurity attacks, e.g., \cite{gillen2020design,salim2024digital}.
    \item Using visualization to debug Slingshot network congestion, i.e., identify hot spots in the system.
\end{itemize}

\noindent All of these use cases directly impact performance, power consumption, and operational efficiency. From our analysis, we identified several categories for potential use cases: augmented reality, forensic analysis and diagnostics, predictive modeling, failure detection, operational optimization, ``what-if'' scenarios, and virtual prototyping. 

\begin{finding}
    From discussions with various stakeholders including HPC engineers and project managers from both data centers and industrial suppliers, we found that there was a significant need to develop a robust comprehensive digital twin framework for data centers. Such a tool would be valuable for forensic diagnostics and augmentations of operational systems as well as virtual prototyping for future systems. 
\end{finding}

The second step in developing the \gls{DT} involved gathering all the required technical specifications covering the various facets of the architecture, such as converter efficiency curves for rectifiers and voltage converters, pump curves, thermal resistance curves of cold plates, etc. Gathering such information involved many challenges, such as contacting several different organizations, working with different file formats, etc. Understanding and documenting the various technical requirements will be helpful for future developments of digital twins, especially if standard exchange formats can be developed for specifying all the necessary information required to build the \gls{DT}. 

\begin{finding}
\gls{DT} development is a holistic effort requiring everyone to be on board, making it challenging to get started; it touches every aspect of the organization, crossing boundaries of the system and organizations. To effectively integrate digital twins with the procurement lifecycle, it is important to identify the stakeholders for each component subgroup to ensure the model is ready as the system comes online. 
\end{finding}

\subsection{Resource Allocator and Power Simulator}
\label{sec:raps}

Our requirements analysis and gathering revealed the need for a module that could accurately simulate resource allocation and power prediction. Therefore, we have developed the \gls{RAPS} module, which is a tight integration of both the job scheduler in concert with dynamic power consumption calculations, which we describe in detail in this section. 
Algorithm \ref{alg:scheduler} shows the pseudocode for the \gls{RAPS} module. An initial array of jobs is created either synthetically or from telemetry data, where each job is characterized by: (1) the number of nodes required, (2) the wall time, and (3) CPU/GPU utilization traces\footnote{Since our system telemetry lacks CPU/GPU utilization, we linearly interpolate power to utilization. In the future, we plan to use profiling to obtain CPU/GPU/network utilization traces for a set of representative applications, as discussed in \cite{holmen2024towards}.} for a given trace quanta.\footnote{Set to 15s in this work to correspond with system telemetry data and will be further discussed in Section \ref{sec:vv}.} Jobs from telemetry may be replayed using the physical twin's scheduling policy, or may use a built-in scheduler, as described in \ref{sec:sched}, which is also used for synthetic jobs. Time is advanced every second and power is computed at every second for the system; however, the cooling model is only called every 15s during the simulation. 

\subsubsection{Estimation of Power Conversion Losses}
\label{sec:loss}

As shown in Fig. \ref{fig:circuit}, each rack of Frontier consists of four shelves, each shelf has two chassis, and each chassis contains four active rectifiers and eight compute blades, yielding a total of 64 blades and 32 rectifiers per rack.
Each rack is directly supplied with three-phase power from the distribution transformer switchboard. The three-phase power is distributed over the rack supplying 32 power rectifiers connected in parallel. The group of four rectifiers shares a common output DC bus, providing power to eight blades; these blades, in turn, feed power to 16 step-down DC-DC converters that are connected in parallel.

The role of the rectifier is to convert the three-phase AC energy supplied from the distribution transformer into constant energy that is distributed among the blades. In every partition, constant energy is distributed through a common DC bus, so that in case of rectifier failure, blades are continuously powered and should perform their job without any interruption. Each blade consists of two isolated DC-DC step-down converters, also known as \glspl{SIVOC}, as shown in Fig. \ref{fig:circuit} \cite{hpe2021_sivoc}, which further steps down the 380V DC voltage from the rectifier into 48V DC voltage that supplies power to the node. The total efficiency of the energy conversion $\eta_{system}$ of the active rectifier and the SIVOC converter is given by:

\begin{equation}
\label{eq:eta}  
\eta_{system} = \eta_{R} \eta_{S} = \frac{P_{R_{DC}}}{P_{R_{AC}}}
\frac{P_{S_{48V}}}{P_{R_{DC}}}= \frac{P_{S_{48V}}}{P_{R_{AC}}}
\end{equation}

\noindent and the total power conversion loss $P_{L}$ is given by:

\begin{equation}
\label{eq:loss}
P_{L} = P_{L_R} + P_{L_S} = P_{R_{DC}}(P_{R_{AC}} - P_{S_{48V}}), 
\end{equation}
\vspace{1mm}

\noindent where $P_{R_{AC}}$ is the rectifier input power, $P_{R_{DC}}$ is the rectifier output power, and $P_{S_{48V}}$ is the SIVOC output power. 
$\eta_R$ and $\eta_S$ is the rectifier and converter efficiency, which were respectively determined to be 0.96 and 0.98. Therefore the total system efficiency according to (\ref{eq:eta}) is roughly\footnote{Note, that these efficiencies are simplifications for this discourse (yet still within one percent of the actual value); in reality, the efficiency slightly varies based on input power, which is discussed in more detail in \cite{wojda2024dynamic}.} 0.94.

\begin{figure*}[t]
\centering
\includegraphics[width=6in]{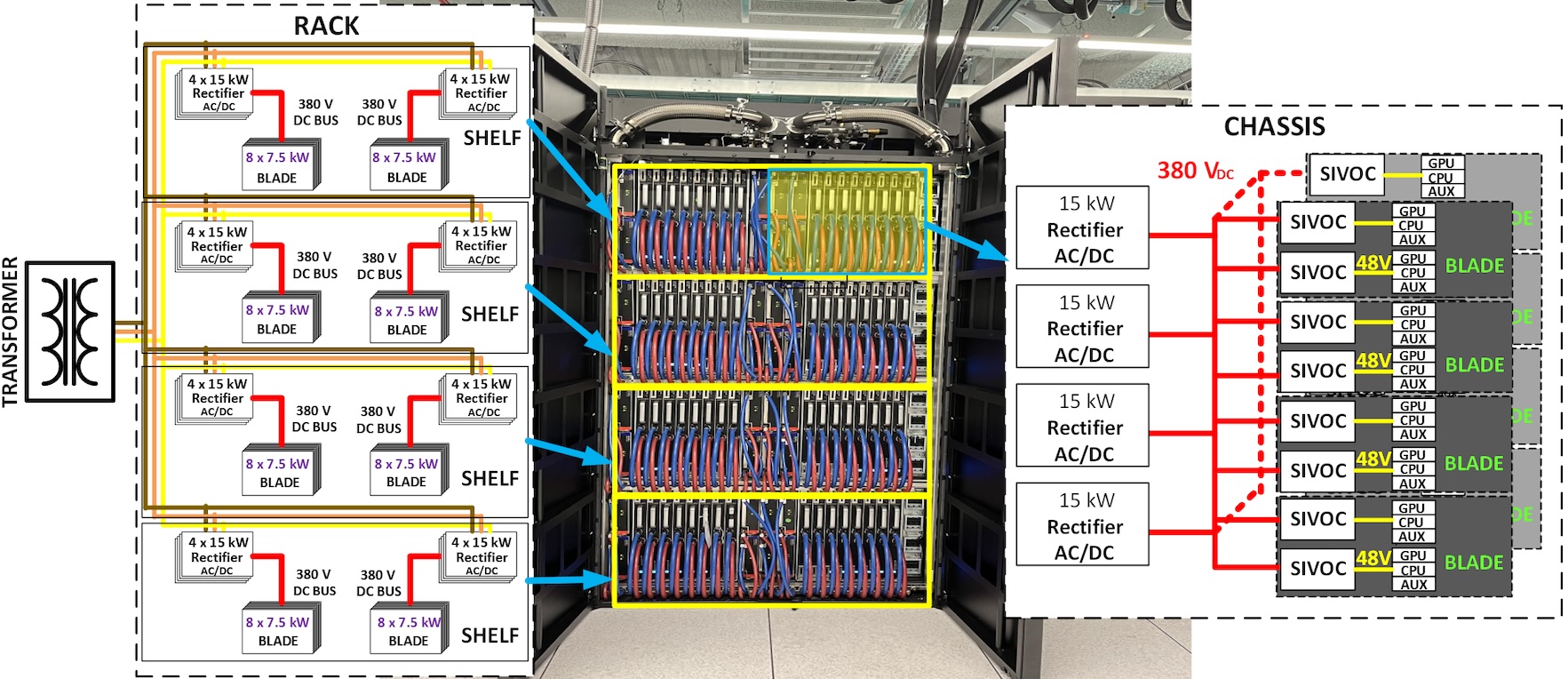} \\
\caption{Frontier rack-level power distribution and voltage conversion.}
\label{fig:circuit}
\end{figure*}

\begin{table}[t]
\centering
\caption{Component Overview of the Frontier Supercomputer}
\label{tab:frontier_specs}
\centering
\hfill
  \begin{tabular}{L{\widthof{Rectifiers per Rack}}R{\widthof{\textbf{Quantity}}}}
        \toprule
        \textbf{Component} &\textbf{Quantity}\\
        \midrule
        Number of CDUs   & 25\\
        Racks per CDU    & 3\\
        Chassis per Rack & 8\\
        Rectifiers per Rack & 32\\
        Blades per Rack   & 64\\
        Nodes per Rack  & 128\\
        SIVOCs per Rack & 128\\
        Switches per Rack & 32\\
        Nodes Total & 9472\\
        \bottomrule
\end{tabular}
\hfill
\centering
  \begin{tabular}{L{\widthof{Switch (Avg)}}R{\widthof{8700 W}}}
        \toprule
        \textbf{Component} & \textbf{Power}\\
        \midrule
        GPU (Idle) & 88 W \\
        GPU (Max)  & 560 W \\
        CPU (Idle) & 90 W \\
        CPU (Max) & 280 W \\
        RAM (Avg) & 74 W \\
        NVMe (Avg) & 30 W \\
        NIC (Avg) & 80 W \\
        Switch (Avg) & 250 W \\
        CDU (Avg) & 8700 W \\
        \bottomrule
\end{tabular}
\hfill
\end{table}

\subsubsection{Dynamic Power Estimation}
Frontier is composed of 9472 ``Bard Peak'' nodes, with each blade housing two such nodes. Each node contains an AMD EPYC\texttrademark\ 7A53 ``Trento'' 64-core 2-GHz CPU and four AMD Instinct MI250X GPUs \cite{atchley2023frontier}. Each node's power can be computed by summing its individual components summarized in Table \ref{tab:frontier_specs}: 

\begin{equation}
    P_{node} = P_{CPU} + 4 P_{GPU} + 4 P_{NIC} + P_{RAM} + 2 P_{NVMe}
\end{equation}

\noindent The [idle, peak] values for $P_{CPU}$ and $P_{GPU}$ respectively are [90, 280] and [88, 560] watts. We set $P_{RAM} = 74$W based on the mean RAM power, $P_{NIC} = 20$W (four per node), and we use $P_{NVMe} = 15$W (two per node). Every second in the simulation, $P_{node}$ is computed for every node by linearly interpolating between [idle, peak] power values for the time-indexed value in the CPU/GPU utilization traces to get the $P_{CPU}$ and $P_{GPU}$ values. After computing $P_{node}$, rectification and conversion losses $P_{L_R}$ and $P_{L_S}$ are applied. Then, power is summed at the rack level, which includes the 250W of power \cite{desensi2020depth} for each of the 32 network switches per rack:

\begin{equation}
    P_{rack} = \sum_i^{N=128} P_{node}^i + 32 P_{switch}
\end{equation}

\noindent Next, the three racks associated with each \gls{CDU} is summed, which gives 25 dynamic power outputs, corresponding to the 25 \glspl{CDU}. The power output is multiplied by a cooling efficiency\footnote{Computed from telemetry data as heat removed divided by power consumed, discussed in Section \ref{sec:vvraps} and plotted in Fig. \ref{fig:hpl}.} of 0.945 before feeding it into the cooling model, which is described in Section \ref{sec:cooling}. 
To get the total system power, $P_{system}$, we sum these 25 power values and add the power cost to operate the pumps in each of the 25 \glspl{CDU}, which is a constant 8.7 kW per \gls{CDU}, or 217.5 kW in total. We calculate the system at peak utilization (9472 nodes, with CPUs and GPUs operating at full capacity) to consume 28.2 MW and show a breakdown of the individual power contributors in Fig. \ref{fig:pie}. 

\subsubsection{Synthetic Workloads}

In order to model system workloads, we simply analyze system telemetry data to obtain average and standard deviations for quantities such as average job arrival time $t_{avg}$, number of nodes required, and wall time. Then it simply generates randomly distributed values for average CPU/GPU utilizations. 
We still have much work to do on the topic of ``application fingerprinting'' to develop more accurate models of jobs. This is an area where AI/ML can be useful for developing a job generator. One promising tool that can be used in this capacity is Kronos \cite{Kronos_ECMWF}. 

\subsubsection{Modeling Job Arrival and Scheduling}
\label{sec:sched}

\textsc{RunSimulation} submits jobs to the queue according to a Poisson process \cite{fan2021deep}, where an exponential distribution is used to model the time between job arrivals given by:

\begin{equation}
    \tau = -\frac{\ln(1 - U)}{\lambda}
\end{equation}

\noindent Here, $\tau$ represents the time interval for the next job submission, determined by a uniformly distributed random variable $U$ in the interval (0, 1); $\lambda$, defined as the inverse of the average arrival time ($t_{avg}$), is a configuration parameter that can be determined from telemetry data as $1 / t_{avg}$, where $t_{avg}$ denotes the average interval between job arrivals.
Jobs are scheduled according to a given policy, such as Shortest Job First (SJF) or First Come First Served (FCFS), with plans to soon implement more sophisticated algorithms and evaluate their impact on the overall system.

\subsubsection{Output Statistics}
At the end of the run, a report is provided that outputs statistics on: (1) the number of jobs completed, (2) the throughput (jobs/hour), (3) average power consumed in MW, (4) total energy consumed in MW-hr, (5) rectification and conversion losses in MW (6) CO2 emissions in metric tons, and (7) total energy costs in USD. CO2 emissions are calculated by multiplying the average system power $\bar{P}_{system}$ by the following emissions factor from \cite{EPA2023}:

\begin{equation}
    E_f = E_I \times 1\,\text{metric ton} / 2204.6 \,\text{lbs} \times 1/\eta_{system}
\end{equation}

\noindent where $E_I$ represents the emission intensity in CO$_2$/MWh, currently set at 852.3; however, this value can vary regionally and even hourly \cite{li2023toward}.

\subsubsection{Deployment}
\label{sec:deployment}
In order to make \gls{RAPS} more accessible as a useful tool for the technical staff and HPC engineers, we deploy it on an internally hosted Kubernetes (K8s) cluster. This allows us to host a dashboard interface, developed in TypeScript with ReactJS, shown in Fig. \ref{fig:interfaces}, for launching simulations to perform ``what-if'' scenarios and plot/analyze the results. Each case runs in a separate K8s pod, and the results of each experiment can be saved in the Apache Druid database and recalled later. This type of deployment enables users to easily perform a wide range of experiments and quickly plot the results. The dashboard also relies on a backend HTTP REST API for accessing telemetry data from the physical twin.

\begin{finding}
There are multiple approaches to accurately modeling power, based on time granularity or level of detail. We found that using a coarse-grained, job-centric simulation -- which traces CPU/GPU utilization and includes resource allocation and power loss modeling -- provides good estimates of dynamic power for an exascale digital twin.
\end{finding}

\begin{figure}[t]
    \centering
    \includegraphics[width=2.7in]{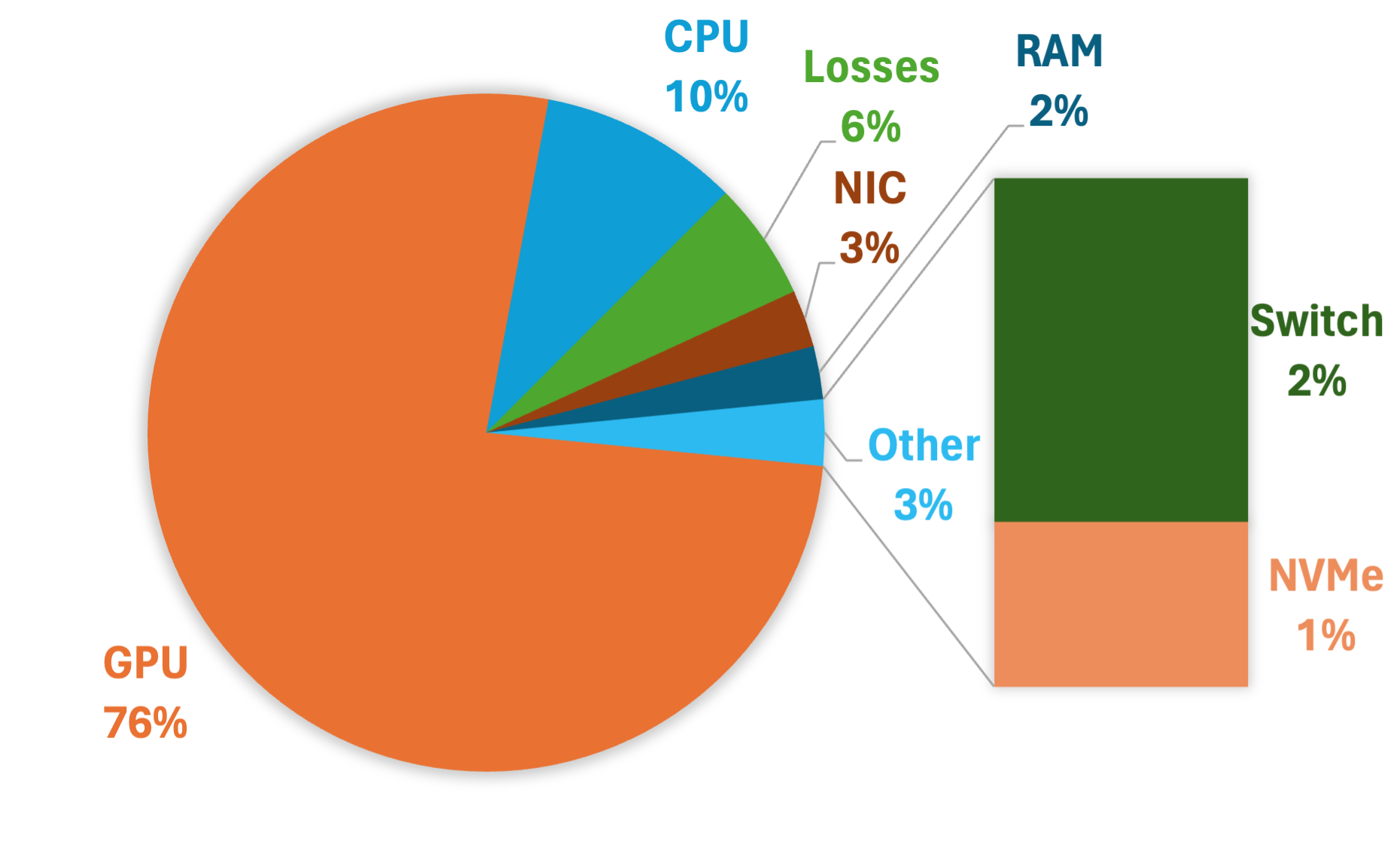} \\
    \caption{Frontier power utilization breakdown based on peak CPU/GPU utilization of its 9472 nodes.}
    \label{fig:pie}
\end{figure}

\begin{algorithm}[t]
\caption{\gls{RAPS} Pseudocode}\label{alg:scheduler}
\begin{algorithmic}[1]
    \Procedure{Main}{}
        \State Initialize scheduler
        \State Generate list of jobs to be executed
        \State Call \textsc{RunSimulation}
    \EndProcedure
    \Procedure{RunSimulation}{jobs, timesteps}
        \For{each timestep}
            \State Add newly arriving jobs to pending queue
            \State Call \textsc{ScheduleJobs} with pending jobs
            \State Call \textsc{Tick}
        \EndFor
    \EndProcedure
    \Procedure{Tick}{}
        \Comment{Called every second}
        \State Increment current time
        \For{each running job}
            \If{job is completed}
                \State Release nodes
                \State Update power state to idle
            \EndIf
        \EndFor
        \State Recalculate power consumption
        \State Apply rectification and conversion losses
        \If{$timestep \bmod 15 = 0$}
            \Comment{Update every 15s}
            \State Call FMU cooling model
            \State Update UI/Status
        \EndIf    \EndProcedure
    \Procedure{ScheduleJobs}{jobs}
        \For{each job in jobs}
            \If{enough nodes available}
                \State Assign nodes to job
                \State Update power state to active
            \Else
                \State Add job to pending queue
            \EndIf
        \EndFor
    \EndProcedure
\end{algorithmic}
\end{algorithm}

\subsection{Cooling Model} 
\label{sec:cooling}

\begin{figure*}[htb]
    \centering
    \includegraphics[width=\textwidth]{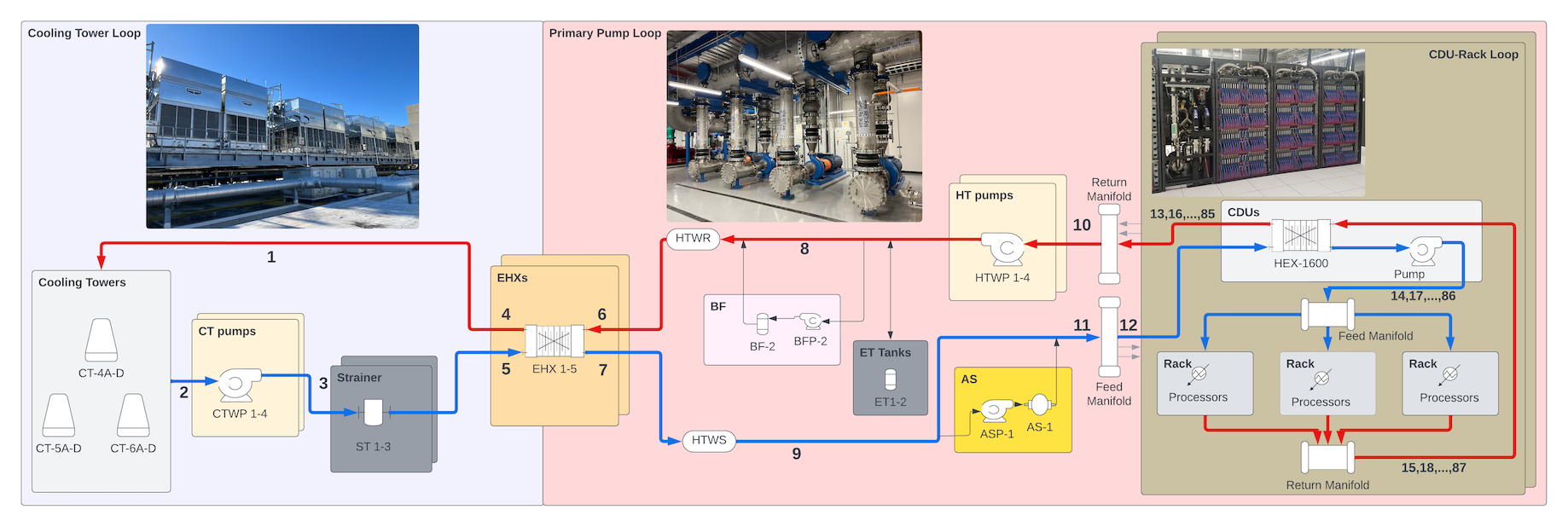}
    \caption{Simplified schematic of Frontier cooling system with enumerated locations where the cooling model predicts pressures, temperatures, and flow rates.}
    \label{fig:cooling_schematic}
\end{figure*}

Having presented the \gls{RAPS} module's ability to predict and manage power consumption, we now transition to modeling Frontier's liquid cooling system, designed to rapidly dissipate the vast amounts of energy lost as heat due to resistive loads caused by the many voltage converters, GPUs, CPUs, and DIMMs in the system.

\subsubsection{Frontier's Cooling System}

Fig. \ref{fig:cooling_schematic} shows a simplified layout of the overall cooling system for Frontier, which contains three cooling loops joined by heat exchangers. 
The \textit{cooling tower loop} circulates through five cooling towers (CT), each with four cells, totaling 20 independent cells. The flow continues through four cooling tower water pumps (CTWP1-4) at approximately 9000-10000 gpm; it then passes through the five intermediate heat exchangers (EHX1-5).
The \textit{primary pump loop} flows from EHX1-5 through the four high temperature water pumps (HTWP1-4) at approximately 5000-6000 gpm. The flow then reaches the HEX-1600 heat exchangers, one per each of the 25 \glspl{CDU}, which serve all 74 racks.
In each of these 25 \textit{\gls{CDU}-rack loops}, flow passes through the HEX-1600 heat exchanger to the \gls{CDU} pump and then the flow is split to serve three racks. In each rack, the flow passes through 64 compute blades (each with two nodes), through two CPU cold plates, and eight GPU cold plates. Therefore, each \gls{CDU} cools a total of 192 blades, or 384 nodes. 

\subsubsection{Modelica}

Rather than using a high-fidelity computational fluid dynamics (CFD) approach, which is typically used in air-cooled systems, we have opted to use Modelica -- a system-level modeling language that offers a good balance between advanced predictive capability and simulation time.
Modelica is an open-source, acausal, object-oriented programming language used for modeling complex cyber-physical systems \cite{modelica}. 
The Modelica standard library can be easily extended using a variety of commercial and open-source packages.
Both commercial and open-source \glspl{IDE} are available to develop and run the Modelica model. 
One of the principal advantages of Modelica lies in its acausal, declarative style which affords a clean separation between the \glspl{ODE} that are being coded by the user, and the solution procedure for the complex system of equations.

\begin{finding}
For modeling the data center cooling, while there are several commercially available tools, they offer limited extensibility while also being cost prohibitive (on the order of tens of thousands of dollars per year). 
Despite its steep learning curve, Modelica as an open-source framework provides the best value and path for meeting our current needs, as well as for building a vibrant digital twin community.
\end{finding}

\subsubsection{Model Dependencies}

The `Modelica.Fluid' library, part of the Modelica Standard Library, solves zero-dimensional and one-dimensional thermo-fluid flow in a variety of fluidic components such as pipes, fluid machines such as pumps, vessels, valves, fittings, etc \cite{casella2006modelica, franke2009}.
The governing equations are formulated based on a finite volume method with a staggered grid scheme for momentum. 
The component equations and the media models are decoupled from each other in the `Modelica.Fluid' library allowing the user to specify incompressible or compressible media, single-component or multi-component mixtures (including two-phases), as well as separate correlations for pressure drop and heat transfer coefficients \cite{casella2006modelica, franke2009}. 
Different closure relations are available based on flow regime, i.e., laminar, turbulent and transition flow. 

For the current study, the \gls{trans} library was utilized in conjunction with the Modelica buildings library (MBL).
\gls{trans} is a Modelica-based open-source library to enable rapid development of dynamic, advanced energy systems with an extensible system modeling tool \cite{greenwood2017a, greenwood2020}.
On the other hand, MBL  provides components for modeling building performance, including HVAC systems, and controls \cite{wetter2014modelica}. 
Specifically, the variable fan speed-based cooling tower model was used from MBL. 
All other components are either derived from or directly accessible within the \gls{trans} library.

\subsubsection{Thermo-Fluids Model}

All the sub-models shown in Fig. \ref{fig:cooling_schematic} are constructed with volumes (reservoirs) for mass sources, resistances for pressure drops, pumps, heat exchangers, and sensors, according to the templated layout described in \cite{greenwood2017b}.
The model takes as inputs wet-bulb (outdoor) temperature and heat extracted in watts for each of the 25 \glspl{CDU}. The model produces a total of 317 outputs for each timestep of simulation (currently 15s), which is broken down as follows. 
For each of the 25 CDUs, there are 11 model outputs: 
work done by the CDU pump (station 14); primary and secondary flow rates (stations 12, 14 respectively); supply and return temperatures and pressures (stations 12-15).
For the primary pump loop, the model outputs information related to the number of pumps and heat exchangers staged, and power consumption and pump speed for each of the four \glspl{HTWP}. For the cooling tower loop, the model outputs the number of \glspl{CT} staged and power consumed by the four \glspl{CTWP}, and the power consumed by the 16 \gls{CT} fans. The \gls{PUE} is computed by summing the total facility energy and dividing by $P_{system}$.  

\subsubsection{Control system model} 

In order to accurately model the cooling system and predict \gls{PUE}, it was necessary to model Frontier's control system, which enables the staging of cooling towers, heat exchangers, and pumps. 
The control system logic is divided between the \gls{CEP} and the supercomputer.
A detailed overview of the control system logic is beyond the scope of this paper; therefore, we provide a concise explanation here.\footnote{See Kumar et al. \cite{kumar2024thermo} for a detailed description of the cooling model.} The Modelica model captures the essentials of the control logic, which activates once the physical cooling system begins auto-operation, after the start-up sequence is complete. The model is described for each cooling loop here:

\begin{itemize}
\item \textit{\gls{CDU}-rack loop}: A \gls{PID} controller is used to regulate the \gls{CDU} relative percent pump speeds based on the loop differential pressure, and a control value is used to regulate the primary coolant flow based on a set secondary supply temperature. Most of the PID parameters have been taken from the physical controller where available, and tuned using telemetry data where parameters were not available. Both the \gls{CDU} pumps are assumed to be in operation at all times with the same speeds -- a reasonable approximation as evidenced by telemetry data. 

\item \textit{Primary pump loop}: A \gls{PID} controller is used to regulate the four \glspl{HTWP}. The \glspl{HTWP} are staged up/down depending on the relative percent pump speeds of the running pumps.
The \glspl{EHX} are staged based on the number of \glspl{CT} in operation. 

\item \textit{Cooling tower loop}: The \gls{CTWP} speed is regulated based on the \gls{CT} supply header pressure, which is maintained within a given pressure range; this informs the staging up/down of the four \glspl{CTWP} in concert with the relative percent speeds of the running pumps. 
The \glspl{CT} are staged up/down based on header pressure and the gradient of the \gls{HTWS} temperature. 
\end{itemize}

Therefore, the criteria to achieve \gls{HTWS} temperature stability informs both the staging of the \glspl{CT} directly and the \glspl{EHX} indirectly.
This non-linearity is handled in the model via a delay transfer function between the primary pump loop,  which requires the number of \glspl{CT}, and the cooling tower loop, which requires the \gls{HTWS} temperature. 
Therefore, any disturbance in a \gls{CDU} would affect its control valve, which in-turn would affect the overall system to respond as just described. 
Future efforts will focus on optimizing the control parameters to achieve better system stability such as responding quickly to a surge in power to the \gls{CDU}. 

\subsubsection{Exporting the Cooling Model} 
Using the \gls{FMI} standard \cite{fmi_standard}, the cooling model is integrated into the digital twin framework as a \gls{FMU}. 
An \gls{FMU} is a model which has been wrapped in the standard \gls{FMI} interface and can therefore be used in any software or deployment scenario which has implemented the \gls{FMI}. In this effort, the Dymola IDE was used to both develop and export the system model as an \gls{FMU}.
The cooling model \gls{FMU} is then imported into the \gls{RAPS} module via the FMPy Python module, which is queried from the visual analytics module via FastAPI \cite{fastapi}.

\begin{finding}
    Using a simplified system-level Modelica model yields good transient solutions. Any additional improvement in fidelity would need to be weighed against model complexity which affects model convergence and simulation time.  
\end{finding}
\FloatBarrier

\subsection{Visual Analytics}
\label{sec:visual_analytics}

After discussing the backend modeling for resource allocation, power consumption, and thermo-fluidic cooling, we discuss how to interface with the models to visualize insights.  
For users, it is important to have effective ways to interact with the \gls{DT}.
This enables the realization of value offered by the combined system, either by providing a good overview of what is happening within the system, or by exploring details that are otherwise only accessible with major efforts.
Visual analytics in the context of digital twins helps users to interact effectively with the \gls{DT} to discover new insights \cite{zheng2023visual}. We primarily explored \textit{\glsfirst{AR}} as a visual analytics tool, but also experimented with the \gls{RAPS} dashboard interface, as discussed in Section \ref{sec:deployment}, for launching simulations and for plotting statistics. 

\begin{figure*}[t]
\centering
\includegraphics[width=\textwidth]{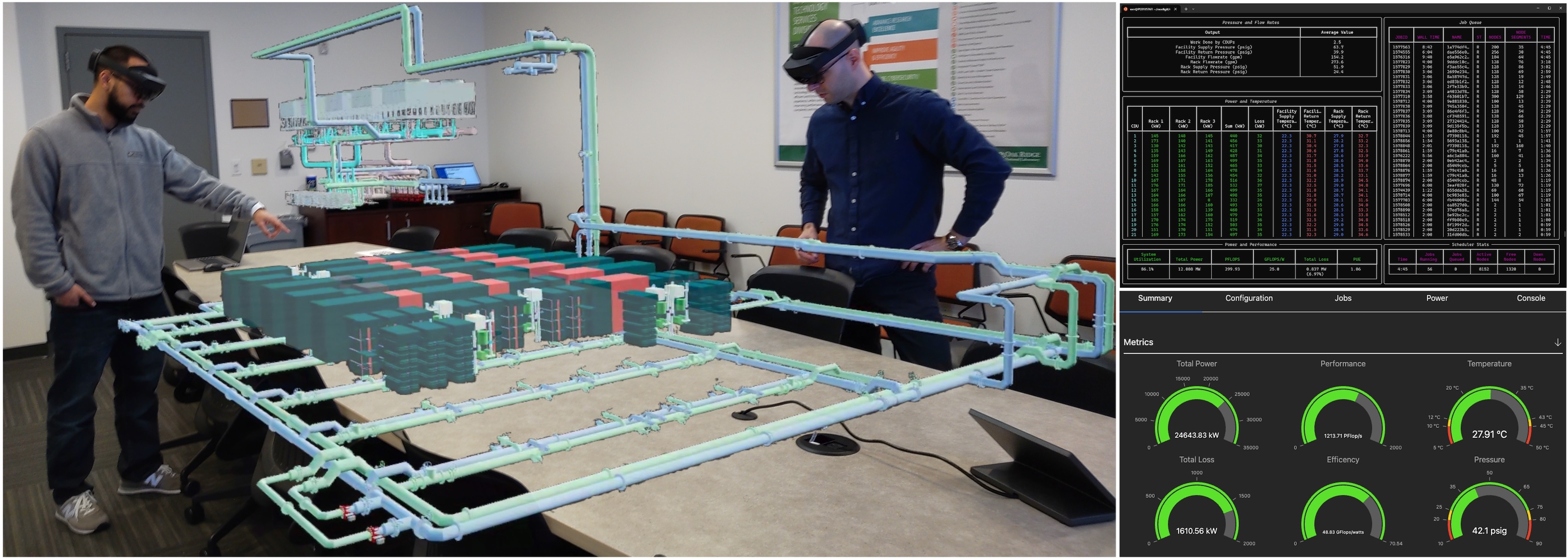}
\caption{\gls{exadigit} interfaces: \gls{AR} model as projected in meeting room (left), terminal interface (top-right), and web-based dashboard (bottom-right).}
\label{fig:interfaces}
\end{figure*}

Representing the physical asset in a 3D virtual space simplifies spatial information correlation more intuitively than using more complex data analysis techniques.
With this in mind, we implemented a virtual representation of Frontier and the \gls{CEP} in \gls{UE5}\footnote{See Maiterth et al. \cite{maiterth2024} for a detailed description of the augmented reality model.} as shown in Figure~\ref{fig:interfaces}, as viewed through a Microsoft HoloLens 2 \gls{AR} headset. 
The visualization allows for system interaction, 3D navigation, and the querying of simulation and telemetry data. 
Users can run this from their desktops or laptops, or as an \gls{AR} application.
Also, in \gls{AR} it is possible to overlay the information onto the physical system, gaining information and insights that are otherwise not realizable. 
An interactive or programmable level of detail was the key to make our \gls{UE5} model performant and responsive with the large number of components and associated telemetry present in the system.

\begin{finding}
For a complex system such as Frontier, with its volume, variety, and velocity (3Vs) of data, we found that augmented reality coupled with dashboards is one of the most effective ways to provide timely visual insights into the system.
This allows us to seamlessly navigate the system complexity and
its simulations, both in space and time, from system overview to blade and component-level views.
\end{finding}

\section{Verification, Validation, \& Functional Tests}
\label{sec:vv}

The National Academy of Sciences, Engineering, and Medicine (NASEM) recently published a report of findings and recommendations for digital twins \cite{nas2023foundational}. One of their key recommendations was to deeply embed verification, validation, and uncertainty quantification (VVUQ) into the development of digital twins. Following such advice, we prioritized extensive V\&V of our power and cooling models after the initial development phase, and also have implemented UQ into our RAPS module.
We see \textit{verification} as running some basic tests to see if the model is performing as expected, whereas \textit{validation} involves comparing the models with more extensive system telemetry data. In this section, we discuss V\&V related to the power and cooling of our \gls{DT}, while also demonstrating telemetry replay for multiple historical scenarios. 

\begin{table}[t]
    \centering
    \caption{Specification of Telemetry Data Used for Validation}
    \label{tab:telemetry_spec}
    \begin{tabular}{@{}ll@{}}

    \textbf{RAPS} & \textbf{Model Schema (Resolution, Length)} \\ \hline
    Inputs & jobs: List[Dict]: \\ 
           & \quad job\_name: str \\ 
           & \quad job\_id: int \\ 
           & \quad node\_count: int \\ 
           & \quad start\_time: float \\ 
           & \quad cpu\_power: List[float] (15s, variable length) \\ 
           & \quad gpu\_power: List[float] (15s, variable length) \\ \hline
    Output & measured\_power: List[float] (1s) \\ \hline 
    
    \vspace{1mm} \\

    \textbf{Cooling Model} & \textbf{Model Schema (Resolution, Length)} \\ \hline
    Inputs & rack\_power: List[float] (15s, 25) \\ 
           & wetbulb\_temperature: float (60s) \\ \hline
    Outputs (CDU) & \{htw,ctw\}\_flow\_rates: List[float] (15s, 25) \\
            & cdu\_temps: List[float] (15s, 25) \\
            & cdu\_pump\_speeds: List[float] (15s, 25) \\
            & cdu\_pump\_power: List[float] (15s, 25) \\ \hline
    Outputs (CEP) & facility\_flow\_rates: List[float] (2m, 2) \\ 
            & \{supply,return\}\_temps: List[float] (1m$\sim$10m, 2) \\ 
            & \{supply,return\}\_pressures: List[float] (30s$\sim$10m, 2) \\ 
            & \{htwp,ctwp\}\_pump\_power: List[float] (10m, 4) \\ 
            & \{htwp,ctwp\}\_pump\_speed: float (2m) \\
            & num\_\{ctwp,htwp,ehx,ct\}\_staging: int (variable) \\
            & pue: float (15s interpolated) \\ \hline
       \end{tabular}
\end{table}

Table \ref{tab:telemetry_spec} shows the telemetry data used to validate the digital twin. This set of telemetry data closely follows what Shin et al. \cite{shin2021revealing} collected and analyzed from the Summit supercomputer, due to the proximity of the operational use cases our work aims to address.

\begin{finding}
We observe that one of the most effective ways to perform verification and validation studies of the power and cooling models is by replaying system telemetry at multiple levels through the digital twin. 
\end{finding}

\subsubsection{Cooling model V\&V} 
\label{sec:vvraps}
A validation study of the entire cooling model, as shown in Fig. \ref{fig:VV_cooling}, was conducted using $~\sim$24 hours of 2024-04-07 telemetry data from the \gls{CEP} and the datacenter down to the level of the \gls{CDU}. 
The only inputs to the model is the power supplied to the 25 \glspl{CDU} (and associated three racks per \gls{CDU}) and the wet-bulb (outside) temperature.
The power supplied to the \glspl{CDU} for the validation exercise was calculated in terms of heat removed by the cooling water: 

\begin{equation}
H = \rho \cdot Q \cdot \Delta T \cdot c
\end{equation}

\noindent where $H$ is the extracted heat  measured in watts, $\rho$ is the density of water in $kg/m^3$, $Q$ is the flow rate in the CDU-rack loop  in $m^3/s$, $\Delta T$ is the temperature differential caused by the heat exchanger in °C and $c$ is the specific heat capacity of water in 
$\text{J} / (\text{kg}\cdot\text{°C})$. 

Comparing model predictions with telemetry data reveals several insights. 
For most parameters, such as those in Fig. \ref{fig:VV_cooling}, the model performs well and is able to predict the response of the physical cooling system to the change in the compute load. 
Overall, both the \gls{RMSE} and the \gls{MAE} of the parameters shown in Fig. \ref{fig:VV_cooling} are within reasonable bounds. 
The model predictions for the \gls{CDU} secondary supply temperatures, although not displayed, exhibit greater fluctuation than the physical system, which does a good job maintaining the temperature at the setpoint. This discrepancy will require further investigation. 

The comparison between the \gls{PUE} predicted by the model and that calculated from telemetry data are shown in Fig. \ref{fig:pue}. 
The model-predicted \gls{PUE} is within 1.4 percent of the telemetry-based \gls{PUE} for the range of data tested. 
Calculations for both the model-predicted and telemetry-based \glspl{PUE} are based on power consumption $P_{\text{AUX}}$ from the following auxiliary systems: CDU pumps, \glspl{HTWP}, \glspl{CTWP} and \gls{CT} fans. Low-power auxiliary systems, such as air-handling systems, were not modeled or included in the \gls{PUE} calculations. 

\begin{figure}[t]
\centering
    \centering
    \subfigure[\centering Primary \glspl{CDU} flow rate predictions (Station 12 in Fig. \ref{fig:cooling_schematic})]{\includegraphics[width=0.22\textwidth, height=0.22\textwidth]{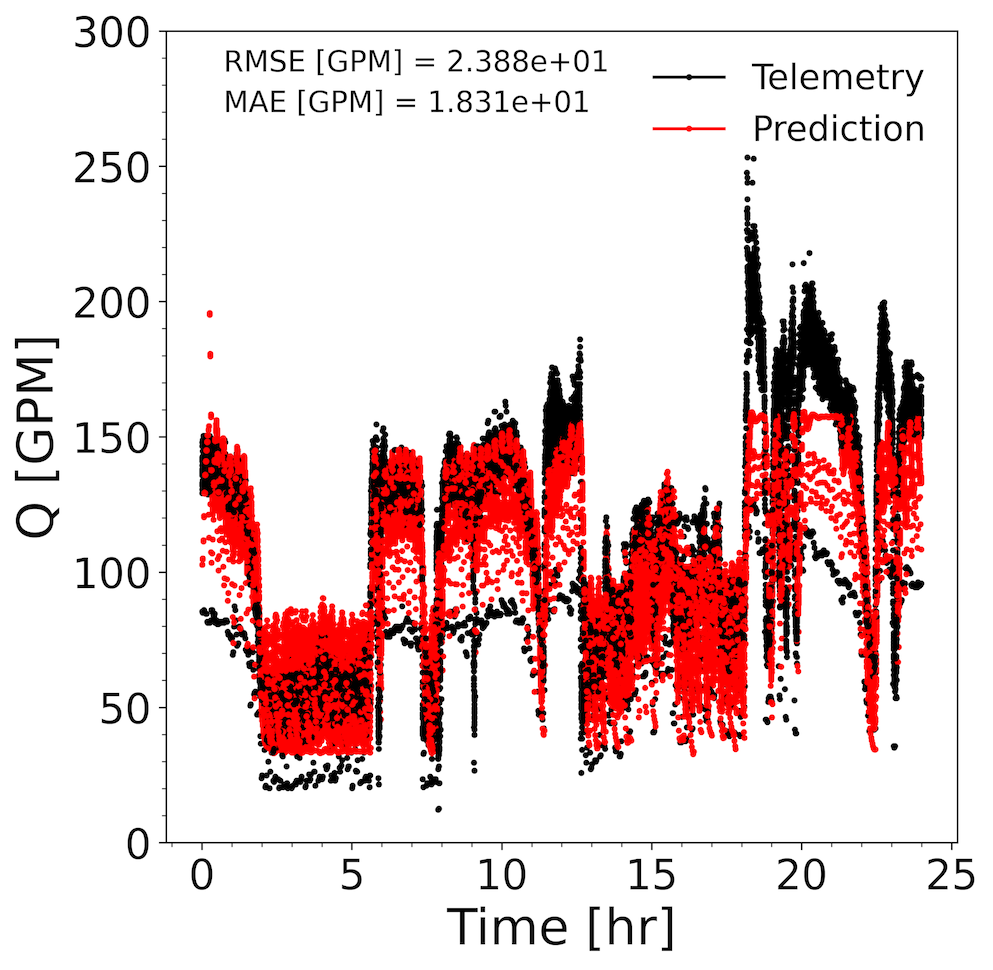}  \label{fig:VV_cdu_pri_Q}} 
    \subfigure[\centering Primary \glspl{CDU} return temp. predictions (Station 12 in Fig. \ref{fig:cooling_schematic})]{ \includegraphics[width=0.22\textwidth, height=0.22\textwidth]{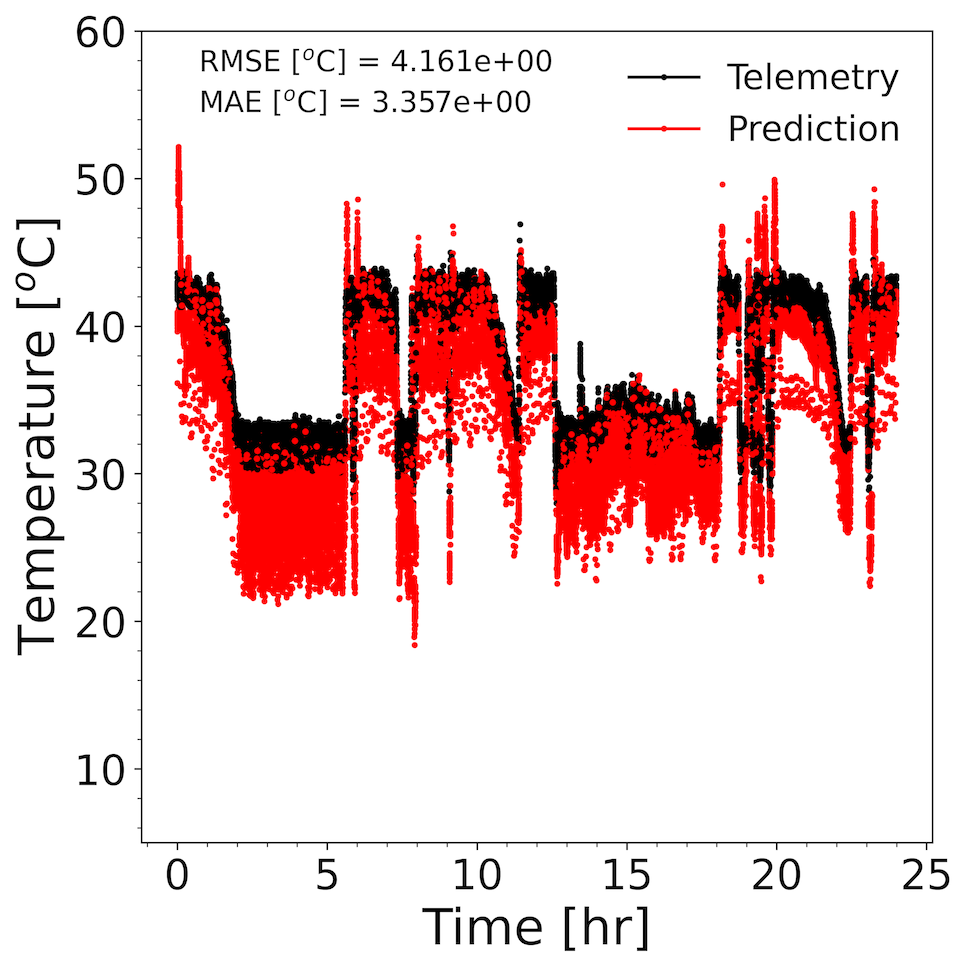}  \label{fig:VV_cdu_pri_Tr}}
    \hfill
    \subfigure[\centering \gls{HTW} supply pressure predictions (Station 10 in Fig. \ref{fig:cooling_schematic})]{\includegraphics[width=0.22\textwidth, height=0.22\textwidth]{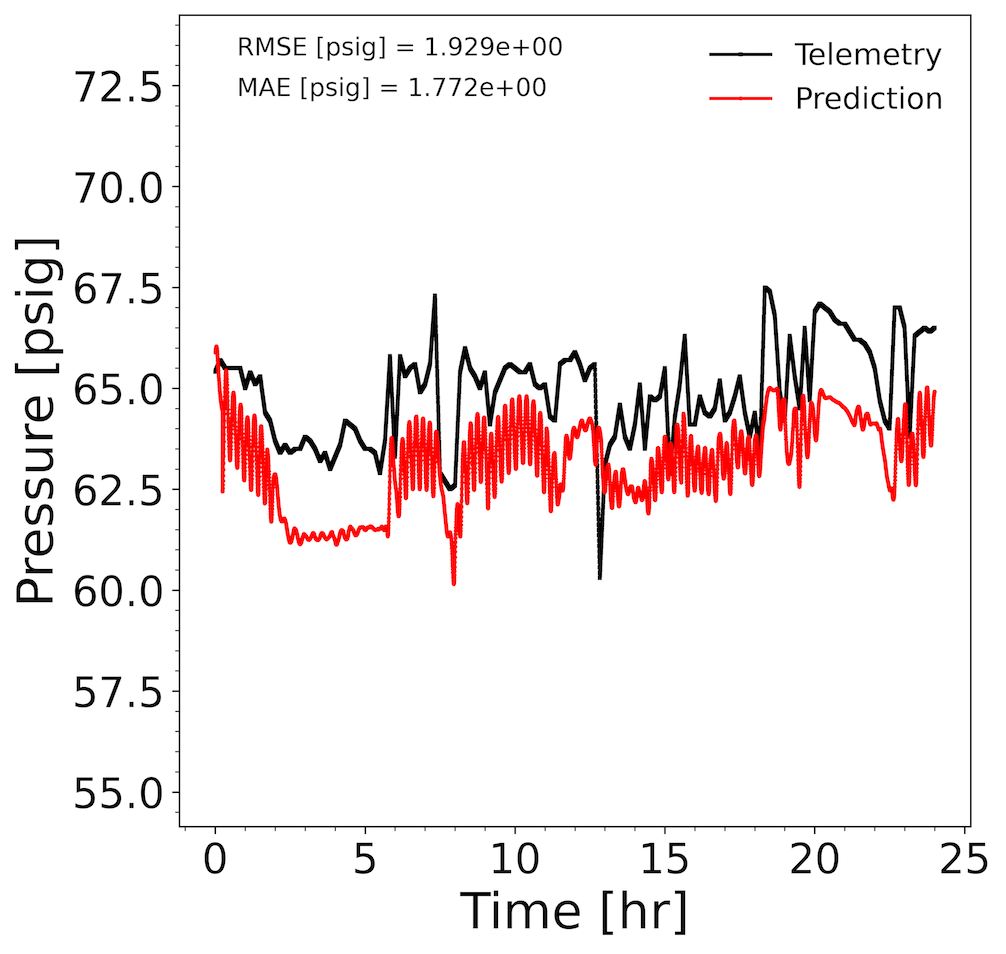}  \label{fig:VV_fac_EHX_pr}}
    \subfigure[\centering PUE Modelica model predictions]{\includegraphics[width=0.22\textwidth, height=0.22\textwidth]{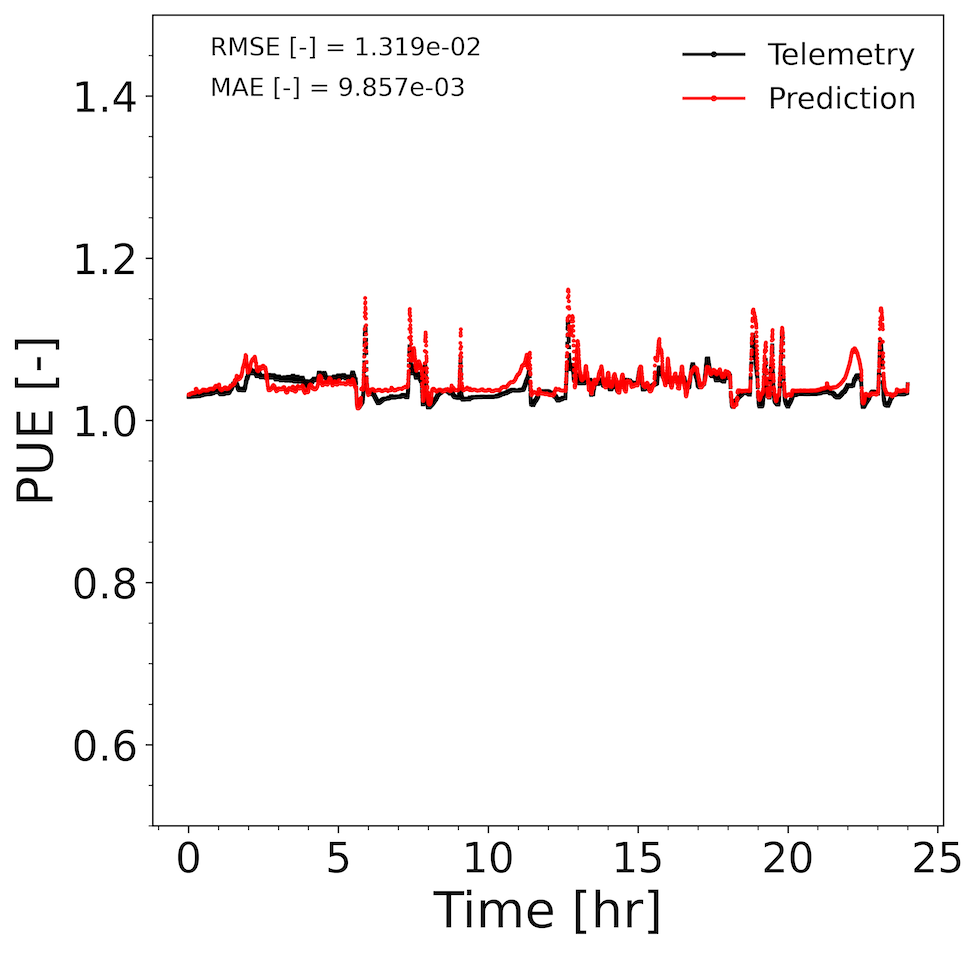} \label{fig:pue}}
    \caption{Cooling model validation tests. Modelica model predictions (exported as an \gls{FMU}) vs. telemetry data for the CDU and the \gls{CEP}.}
    \vspace{-2mm}
    \label{fig:VV_cooling}
\end{figure}

\subsubsection{\gls{RAPS} V\&V} We performed initial verification of \gls{RAPS} by predicting idle and peak power, and also included a \gls{HPL}~\cite{dongarra2003linpack} benchmark.  To set the system at idle, we simply set CPU and GPU utilizations to zero on all nodes; to test the \gls{HPL} core phase, we set all four GPUs on each node to 79\% utilization and the CPU to 33\% utilization (inferred from telemetry data); to test peak power, we set all CPU and GPU utilizations to 100\%.  Table \ref{tab:v_and_v_tests_raps} shows the results of this study.  Fig.~\ref{fig:synthetic} shows synthetic benchmark tests including \gls{HPL} and OpenMxP~\cite{lu2023openmxp} in conjunction with the cooling system predictions of primary return temperature.

\begin{table}[b]
\centering
\caption{\gls{RAPS} power verification tests}
\label{tab:v_and_v_tests_raps}
\begin{tabular}{lcccc}
\hline
\textbf{Tests} & \textbf{Nodes} & \textbf{Telemetry (MW)} & \textbf{\gls{RAPS} (MW)} & \textbf{\% Error} \\
\hline
Idle power       & 9472 & 7.4 & 7.24 & 2.1\% \\
HPL (core) & 9216 & 21.3 & 22.3 & 4.7\% \\
Peak power       & 9472 & 27.4 & 28.2 & 3.1\% \\
\hline
\end{tabular}
\end{table}

\subsubsection{Functional Tests}
After verifying the model could predict idle, HPL core phase, and peak power well, we used \gls{RAPS} to replay 183 days of telemetry data from Frontier, from 2023-09-06 to 2024-03-18. Results are summarized in Table \ref{tab:replay}. Each 24-hour replay takes about nine minutes to run with cooling, or just three minutes without; the entire analysis takes about an hour when running the different days in parallel on a single Frontier node. Average conversion losses amount to 1.14 MW (6.74\%) which works out to about \$900k per year.

Fig.~\ref{fig:hpl} shows a 24-hour period replay, which includes 1238 jobs in total, 400 of which are single-node jobs, and four back-to-back \gls{HPL} 9216-node jobs, among others.
This plot illustrates the instantaneous system power, \(P_{system}\), with both predicted values (in red) and measured values (in black); the energy efficiency, \(\eta_{system}\) (green), as defined in (\ref{eq:eta}); the cooling efficiency (blue), defined as \(\eta_{cooling} = H/P_{system}\); and the system utilization (orange), calculated as the ratio of active nodes to the total available nodes.

Now we can begin to envision ways to improve overall efficiency through virtual modifications to Frontier's \gls{DT}.
The first idea we tested was modifying Frontier to use ``smart load-sharing rectifiers''. The rectifiers reach an optimal efficiency of 96.3\% at 7.5 kW, but near idle the efficiency drops 1-2\%. Instead of sharing the chassis load across all four rectifiers, rectifiers are dynamically staged on as needed, so that rectifiers are always operating at their peak efficiency regions. 
While this modification yielded only a modest efficiency gain of 0.1\%, it translates into a not so insignificant yearly cost savings of approximately \$120k, based on the same 183 days of data in Table \ref{tab:replay}. 
A second test, inspired by \cite{ton2008dc} and \cite{hikari-supercomputer}, focused on switching the Frontier \gls{DT} to direct 380V DC power, instead of AC power. This modification substantially increased the system efficiency from 93.3\% to 97.3\%, a potential savings of \$542k per year, while also reducing the carbon footprint by 8.2\%. 

\begin{figure}[t]
\centering
\includegraphics[width=3.3in]{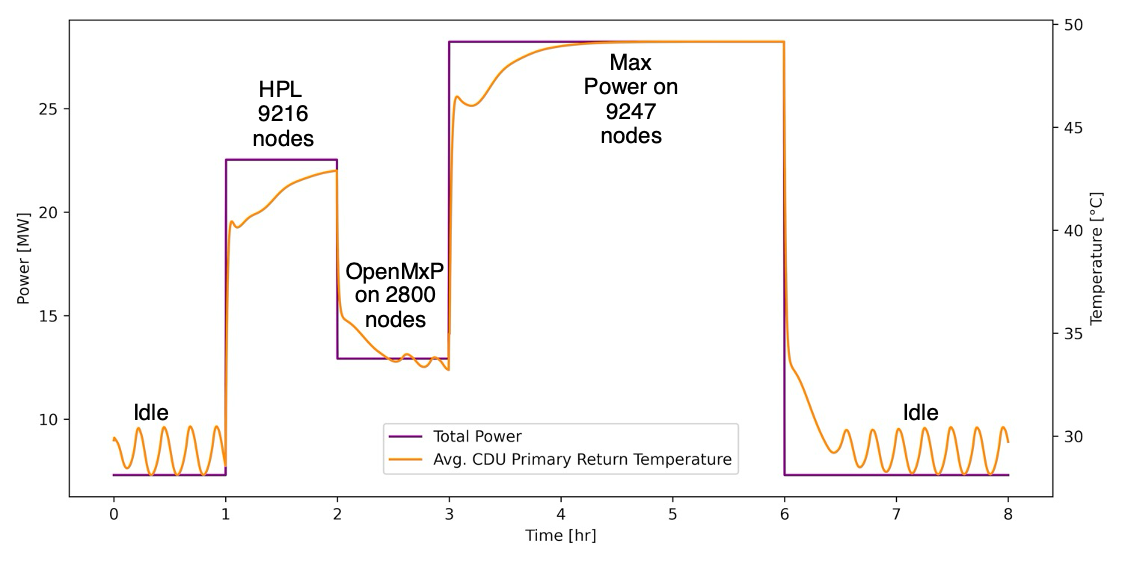} 
\caption{Synthetic benchmark verification test. Total system power predicted by \gls{RAPS} and the transient temperature response predicted by the cooling model.}
\label{fig:synthetic}
\end{figure}

\begin{figure}
\centering
\includegraphics[width=3.3in]{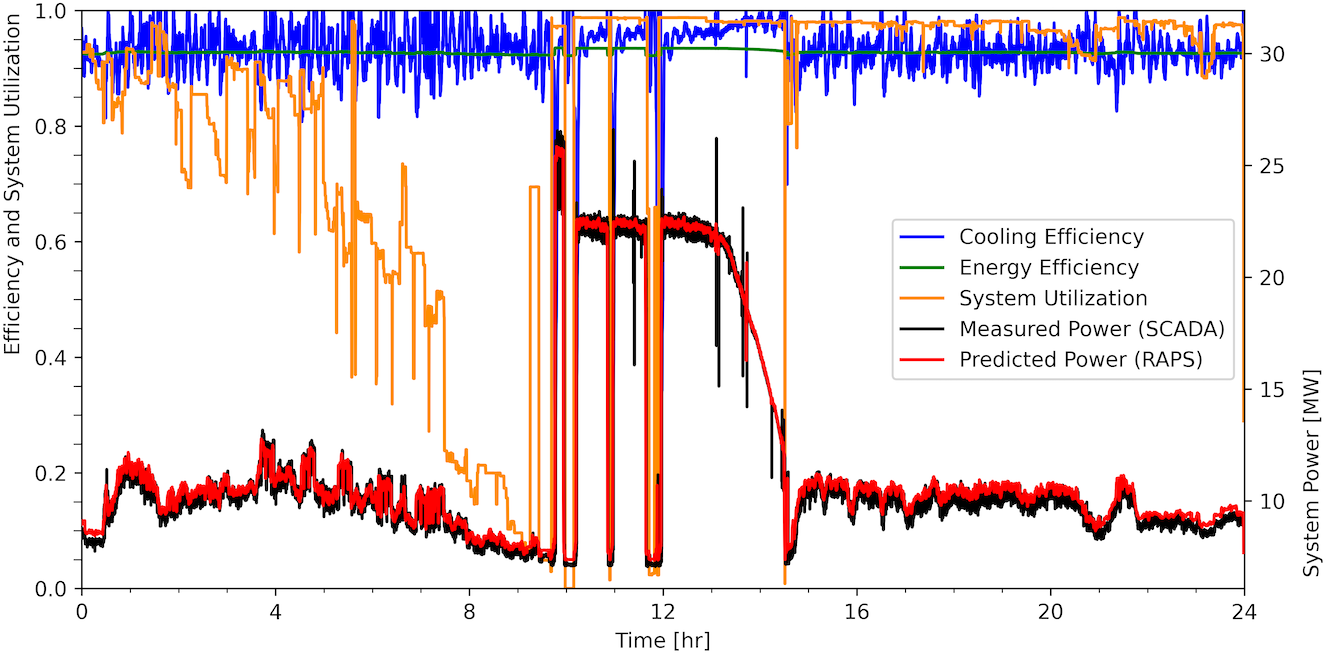} 
\caption{Telemetry replay validation test of 24-hour period on 2024-01-18 for Frontier containing an HPL run.}
\label{fig:hpl}
\end{figure}

\begin{table}[t]
\centering
\caption{Daily statistics of \gls{DT} from telemetry replay of 183 days.}
\begin{tabular}{lllll}
    \toprule
    \textbf{Parameter} & \textbf{Min} & \textbf{Avg} & \textbf{Max} & \textbf{Std} \\
    \midrule
    Avg Arrival Rate, $t_{avg} (s)$ & 17 & 138 & 2988 & 331 \\
    Avg Nodes per Job & 39 & 268 & 5441 & 626 \\
    Avg Runtime (m) & 17 & 39 & 101 & 14 \\
    Jobs Completed & 32 & 1575 & 5157 & 1171 \\
    Throughput (jobs/hr) & 1.3 & 66 & 215 & 49 \\
    Avg Power (MW) & 10.2 & 16.9 & 23.0 & 2.4 \\
    Loss (MW) & 0.52 & 1.14 & 1.84 & 0.15 \\
    Loss (\%) & 6.26 & 6.74 & 8.36 & 0.11 \\
    Total Energy Consumed (MW-hr) & 129 & 405 & 553 & 64 \\
    Carbon Emissions (tons CO\textsubscript{2}) & 53 & 168 & 229 & 26 \\
    \bottomrule
\end{tabular}
\label{tab:replay}
\end{table}

\begin{finding}
Considerable losses are incurred during both AC-DC rectification and DC-DC voltage conversion. Frontier exhibits an average conversion loss of 1.1 MW and maximum of 1.8 MW. Such losses must be accounted for in digital twins, especially when using such a tool to study energy efficiency.
\end{finding}

\section{Generalizing \gls{exadigit}}

While we initially developed \gls{exadigit} to model Frontier, we later worked on generalizing the framework to be able to support development of a variety of architectures. To this end, we determined to use a number of JSON files for input specification, to minimize the level of code changes that must be made to model a particular system. For example, the generalized version of \gls{RAPS} inputs configuration files describing the system architecture, the cooling system, the scheduler, and the power system. A pluggable architecture was developed for reading different types of bespoke telemetry datasets. We are still actively working to extend \gls{RAPS} to handle multi-partition systems, such as Setonix, which have separate partitions for CPU-only nodes and CPU+GPU nodes. A secondary challenge for generalization is addressing shared node configurations, where each node may be shared by multiple users. This generalized approach has recently been used by others \cite{kabir2024} to model Italy's Marconi100 supercomputer, along with an associated PM100 open telemetry dataset \cite{antici2023pm100}. 

For the thermo-fluids model, an automated cooling system model (AutoCSM) method was developed that automates much of the process of developing cooling systems for digital twins \cite{greenwood2024thermo}. AutoCSM, based on Python, inputs a JSON input specification of the architecture of the system, and outputs an initial model of the system, which can then be compiled into an \gls{FMU}. Currently, AutoCSM outputs Modelica code, but in the future can be extended to support other system modeling languages such as JuliaSim \cite{rackauckas2022composing}. We have plans to use the AutoCSM approach to model Marconi100's cooling system \cite{ardebili2022rule}. 

Our augmented reality model has been used by others to develop \gls{AR} models for both Finland's LUMI and Australia's Setonix supercomputers \cite{dykes2024}. To simplify the development process, we also plan to extend the code to support dynamic asset generation based on JSON configuration files. Furthermore, we plan on using an OpenXR based implementation to be able to support a variety of head-mounted displays, such as MetaQuest and Apple Vision Pro.
\section{Conclusions \& Future Work}
\label{sec:conclusions}

We presented the development of a comprehensive digital twin framework for modeling liquid-cooled supercomputers called Exa\-DigiT,\footnote{Available at https://exadigit.github.io.} comprising three main modules: a \glsfirst{RAPS}, a thermo-fluidic cooling model, and an AR-based visual analytics module. An extensive amount of work went into modeling the various aspects of the supercomputer and central energy plant.
\gls{exadigit}'s development revealed several important insights: (1) simulations provide the foundational building blocks \glspl{DT}, with machine learning being important for workload characterization; (2) in modeling power and cooling, it was important to strike a balance between fidelity and complexity, achieved through job-centric power simulations and system-level thermo-fluid simulations; (3) augmented reality is an effective means of interactively visualizing the complexity of the digital twin; (4) system telemetry replay is vital for model verification and validation to evaluate the trustworthiness of \glspl{DT}; (5) modeling the up to 1.8 MW of power losses due to rectification and voltage conversion will be key to using the \gls{DT} for energy efficiency studies.

By considering a comprehensive model of the entire supercomputer, we can study complex cross-disciplinary transient behaviors to provide insights into operational strategies, ``what-if'' scenarios, system diagnostics, as well as serving as a design tool for virtual prototyping.
The development of an open-source framework for comprehensive modeling of digital twins has significant implications in multiple areas, particularly for the design of future energy-efficient supercomputers. 
There has been considerable interest among supercomputing centers around the world in both using and contributing to our \gls{exadigit} framework. To accommodate the growing interest in such a tool, we have organized an \gls{exadigit} community, which currently has 93 global participants. This group meets monthly with workgroups meeting regularly to advance the framework's development, use case studies, and its application across various supercomputer architectures. 

While the digital twin framework presented in this paper has focused on asset visualization (L1), telemetry (L2), and modeling and simulation (L4), we aim to focus future efforts on developing data-driven models (L3), such as machine-learned models of workloads, and reinforcement learning agents (L5) that provide continuous feedback to the physical twin. Such agents will enable the creation of a ``living'' \gls{DT} \cite{sarkar2023dcrl, allen2021digital} that operates alongside the physical twin. 
We hope our efforts will stimulate and enable more energy-efficient supercomputers in the future, by providing a powerful, comprehensive framework along with fostering a global community of researchers to work together toward building a strong future for sustainable energy-efficient supercomputing. In the long term, investigating multi-scale approaches to bridge discrete event simulators, such as Gem5 \cite{lowe2020gem5}, SST/Macro \cite{janssen2010simulator}, and SimGrid \cite{camus2018co}, with comprehensive digital twins for end-to-end optimization will be key for advancing the state of the art \cite{suter2024}. 

\section*{Acknowledgments}

This research was sponsored by and used resources of the Oak Ridge Leadership Computing Facility (OLCF), which is a DOE Office of Science User Facility at the Oak Ridge National Laboratory (ORNL) supported by the U.S. Department of Energy under Contract No. DE-AC05-00OR22725. We thank the technical staff at ORNL, without whom this work would not have been possible, including: Scott Atchley, Matt Sieger, Chris Zimmer, Paul Abston, Jim Rogers, Matt Ezell, Robert Gillen, Dane de Wet, Kazi Asifuzzaman, Nathan Parkison, Corey Spradlin, Brian Reagan, John Holmen, Amir Shehata, Nick Hagerty, Seung-Hwan Lim, Ahmad Maroof Karimi. From HPE, we would like to thank Cullen Bash, Tim Dykes, Matt Slaby, and Justin Queen who provided us with invaluable help along the way. Thanks to Jake Webb from Cadre5, LLC for his significant contributions to the dashboard development. Moreover, we want to acknowledge our growing \gls{exadigit} open source community for their enthusiastic support and engagement, which has been a significant source of inspiration and motivation for this work. Finally, ChatGPT was utilized for converting Python code into pseudocode in Algorithm \ref{alg:scheduler}, grammar enhancements, and assisting with table formatting.

\bibliographystyle{IEEEtran}
\bibliography{main}

\end{document}